\documentclass[runningheads]{llncs}
\usepackage[T1]{fontenc}
\usepackage{graphicx}
\usepackage{amsmath}
\usepackage{booktabs} 
\usepackage{array} 
\usepackage{multirow}
\usepackage{booktabs}
\usepackage{xcolor}
\usepackage{colortbl}
\definecolor{best}{RGB}{198,224,180}  
\usepackage{cite}
\begin{document}
\title{Toward Vision Language Model-based Assessment of Clinical Quality and Usability of LGE-MR Images for Cardiac Ablation Planning}

\titlerunning{VLM-based IQA of LGE-MRI for Cardiac Ablation Planning}

\author{}
%
%
\institute{}
\author{Bipasha Kundu \inst{1}, Abhishek Chaturvedi\inst{3}, Axel W. E. Wismueller\inst{3}, Richard Simon\inst{2}, \and Cristian A. Linte\inst{1,2} }
\authorrunning{B. Kundu et al.}

\institute{Carlson Centre of Imaging Sceince, RIT, Rochester, NY, USA \and
Biomedical Engineering, RIT, Rochester, NY, USA \and 
Department of Imaging Sciences, University of Rochester Medical Center, NY, USA\\
\email{\{bk7944,calbme\}@rit.edu}}

\maketitle              
\vspace{-2.8em} 
\begin{abstract}
Late gadolinium-enhanced (LGE) cardiac MRI is widely used for left atrial (LA) fibrosis assessment and ablation planning in atrial fibrillation patients as knowledge of fibrotic tissue regions identified from LGE-MRI is critical for catheter ablation. Often times, poor quality images used during ablation planning can cause mislocalization of ablation targets, therefore directly impacting procedure safety and outcome. The decision of whether a scan meets the minimum quality threshold for ablation planning is currently made informally by the reviewing radiologist and is not captured by any automated system, yet it is arguably the most safety-critical output of the image quality assessment (IQA) process. However, variations in image quality caused by noise, motion artifacts, and poor boundary definition significantly compromise the reliability of downstream segmentation and clinical decision-making tasks. Manual quality assessment by expert radiologists is subjective and difficult to scale, while existing automated methods produce scalar scores without interpretable clinical reasoning. In this work, we propose a two-stage vision language model (VLM) framework for clinically grounded image quality assessment of left atrial LGE-MRI. In the first stage, a fine-tuned VLM generates structured radiology-style quality reports predicting five radiologist-defined criteria: Noise, Motion Artifact, LA Boundary Accuracy, PV Region Accuracy, and Under-segmentation Severity. In the second stage, a GPT-based reasoning module maps the predicted quality and reports to a structured quality scores and binary clinical usability decision for ablation planning. We curate a dataset of 60 annotated image slice-text pairs from 20 patients and benchmark four state-of-the-art VLM architectures. InternVL2 achieves the highest criterion-level accuracy (Avg ACC=0.65, PLCC=0.79), while DeepSeek achieves perfect clinical usability agreement (Acc=1.00, $kappa$=1.00). These results demonstrate the feasibility of using VLM based automated quality control for clinical workflows that entail the use of MR images.

\vspace{-.5em}
\end{abstract}

\section{Introduction}
\vspace{-.5em}
The emergence of Large Language Models (LLMs) has marked a paradigm shift in the area of natural language processing. Models such as GPT, LLama, and Gemini~\cite{achiam2023gpt, touvron2023llama, team2023gemini} have demonstrated excellent capabilities in generating human-like text across a wide range of tasks such as natural language reasoning, question answering, and structured text generation~\cite{singhal2023large}. However, despite these advances, LLMs are inherently limited to the textual data and cannot directly process the visual data.

To address this gap, recent research has focussed on extending LLMs into the visual domain through Vision Language Models (VLMs) by jointly encoding the visual and textual information with their strong vision encoder and language model backbones. Most popular VLMs such as LLaVA~\cite{liu2023visual}, DeepSeek~\cite{lu2024deepseek}, Qwen~\cite{hui2024qwen2}, and InternVL~\cite{wang2025internvl3} can generate structured textual responses, commonly knows as human-interpretative responses of visual content, which bridges the gap between perception and semantic understanding.

The medical domain is well positioned to benefit from such VLM based approaches. Recent studies have demonstrated that VLMs can generate clinically relevant reports, answers visual questions, identify different organs, and assist in disease diagnosis based on X-ray, CT and ultrasound imaging exams~\cite{iftee2024organ, hartsock2024vlmreview, yildirim2024multimodal, vavekanand2026smiles, vavekanand2026comprehensive}. These studies suggest that VLMs has the potential to serve as a complement to expert radiologists, particularly where annotation capacity is limited and expensive.

While LLM-VLMs have been explored across various medical imaging tasks, their application towards image quality assessment (IQA) is very limited. IQA plays a very important role in downstream image processing tasks including segmentation, registration, and clinical diagnosis~\cite{illimoottil2023recent, kundu2025assessing, kundu2025investigating, kundu2026motion} by aiming to approximate human perception of image quality. A clear and specific example is the crucial role of cardiac MRI as part of cardiac ablation therapy workflows, where LGE-MRI is widely used to guide subsequent catheter ablation procedure planning by enabling the segmentation, localization and visualization of the scar tissue regions relative to the left atrial geometry and the pulmonary vein anatomy. As such, the accuracy of the subsequent  ablation therapy planning workflow hinges upon the quality of these images, and poor image quality can lead to unreliable delineation of anatomical structures and subsequently potentially sub-optimal treatment planning~\cite{campello2020dlreview}.

In clinical practice, the quality assessment of MRI is typically performed by expert radiologists prior to performing quantitative analysis, but this process is subjective, time consuming and difficult to scale. There are also several quantitative methods for IQA in MRI imaging~\cite{chow2016review}, but most of them rely on reference images that may not exist. Furthermore, these methods use objective metrics such as peak-signal-to-noise-ratio (PSNR), contrast-to-noise-ratio (CNR) or root-mean-square-error (RMSE) which are not consistent with the subjective evaluation~\cite{orkild2025image}. Factors such as anatomical visibility, signal noise, contrast, and overall interpretability influence the clinical decision. This process is subjective and time-consuming, and currently there exists no scalable, automated system that reproduces expert-style quality assessment for cardiac MRI that can assess multiple quality dimensions together and produce an interpretable, actionable clinical output.

VLMs offer a promising solution to these limitations. Unlike scalar regression or objective metric-based approaches, VLMs can generate structured, human-interpretable quality reports describing multiple clinically relevant criteria simultaneously, including noise severity, motion artifact extent, boundary clarity, and pulmonary vein visibility. Critically, ablation planning requires not just criterion-level scores, but a definitive binary usability judgment with respect to whether a scan meets the minimum quality threshold to support reliable LA segmentation and ablation target identification. This decision cannot be derived from any single criterion in isolation — an image featuring acceptable noise, but poor boundary definition remains unsuitable, while an image that features elevated noise, but intact boundary and PV accuracy may still support planning. A secondary LLM reasoning layer can map predicted multi-criteria reports to the desired binary decision, explicitly linking image quality to downstream ablation planning in a clinically conservative and interpretable way. The closest prior work by Orkild et al.~\cite{orkild2025image} proposed automated IQA for LGE-MRI in atrial fibrillation patients using traditional scalar metrics. However, it neither generates interpretable multi-criteria quality reports nor produces an automated binary usability decision for ablation planning. Despite their envisioned potential, no prior work has demonstrated that VLM-predicted quality attributes can be reliably translated into clinically grounded usability decisions~\cite{hartsock2024vlmreview}, and hence these decisions remain informal and un-automated in current clinical practice.

In this work, we propose a two-stage VLM-based framework for clinically grounded image quality assessment of left atrial LGE-MRI images. In the first stage, a fine-tuned VLM generates a structured radiology-style quality report that predicts five criteria: Noise, Motion Artifact, LA Boundary Accuracy, PV Region Accuracy, and Under-segmentation Severity. In the second stage, a GPT-based reasoning module maps the predicted quality and report to a structured score and binary clinical usability decision for ablation planning. Our contributions are as follows: (1) We curate a set of 60 MRI image quality assessment datasets annotated across five criteria by expert radiologists (2) We introduce a two-stage VLM-LLM framework that generates structured radiology-style quality reports and maps them to binary clinical usability decisions; (3) We benchmark four state-of-the-art (SOTA) VLMs under parameter-efficient fine-tuning (PEFT), evaluate using ordinal scores and clinical usability metrics, and demonstrating proof-of-concept feasibility for automated cardiac MRI quality control.

\section{Dataset and Annotation}
\vspace{-.5em}
\subsubsection{Dataset Preparation. }We construct a left atrial LGE-MRI image quality assessment dataset by selecting image samples from the LAScarQS 2022 (Task 1) dataset to be subsequently annotated by expert radiologist~\cite{li2022atrialjsqnet, li2022medical}. To adapt this subjectively annotated MRI data with the VLM and IQA task, we created a image slice-text pair based on the annotated dataset. To reduce the redundancy and ensure anatomical relevance, we selected a fixed number of slices from the central region of each 3D volume where the LA is prominently visible. Specifically, we extracted 3 middle slices around the LA based on anatomical prominence per patient, ensuring maximum visibility of the anatomical structure based on slice index ordering. This strategy focuses the model on clinically informative regions while minimizing the inclusion of non-informative slices (basal or apical regions that feature minimal LA visibility). To prevent data leakage, all train-test slices were split at the patient level. The final dataset consists of 60 annotated MRI slices from 20 patients, representing diverse image quality variations.

While the scale of the current annotated dataset is somewhat limited, it is consistent with other expert-annotated cardiac MRI quality benchmarks reported in the literature~\cite{orkild2025image, piccini2020deep}. Moreover, the study conducted to date and reported here is intended as a proof-of-concept, and future work will expand the validation on larger multi-center expert-annotated cohorts.

\begin{table} 
\centering
\caption{Definition of quality assessment criteria used in this study.}
\label{tab:criteria_definition}
\scriptsize
\begin{tabular}{p{2.2cm} p{7cm} >{\centering\arraybackslash}p{1.8cm}}
\toprule
\textbf{Criterion} & \textbf{Meaning} & \textbf{Score Range} \\
\midrule
Noise 
& Level of graininess or signal disturbance affecting visibility of the left atrial wall
& 0--3 \\
Motion Artifact
& Presence of motion-induced blurring or misalignment affecting the left atrial wall
& 0--3 \\
LA Boundary\\ Accuracy
& Clarity and continuity of the left atrial wall boundary
& 0--3 \\
&  \\
PV Region\\ Accuracy
&  Correct delineation of pulmonary vein ostia
& 0--3 \\
Under-\\segmentation Severity 
& Extent of missing or incomplete representation of LA wall
& 0--3 \\

\bottomrule
\end{tabular}
\end{table}
\vspace{-2.5em}
\subsubsection{Annotation Protocol. }
\begin{figure}
\includegraphics[width=\textwidth]{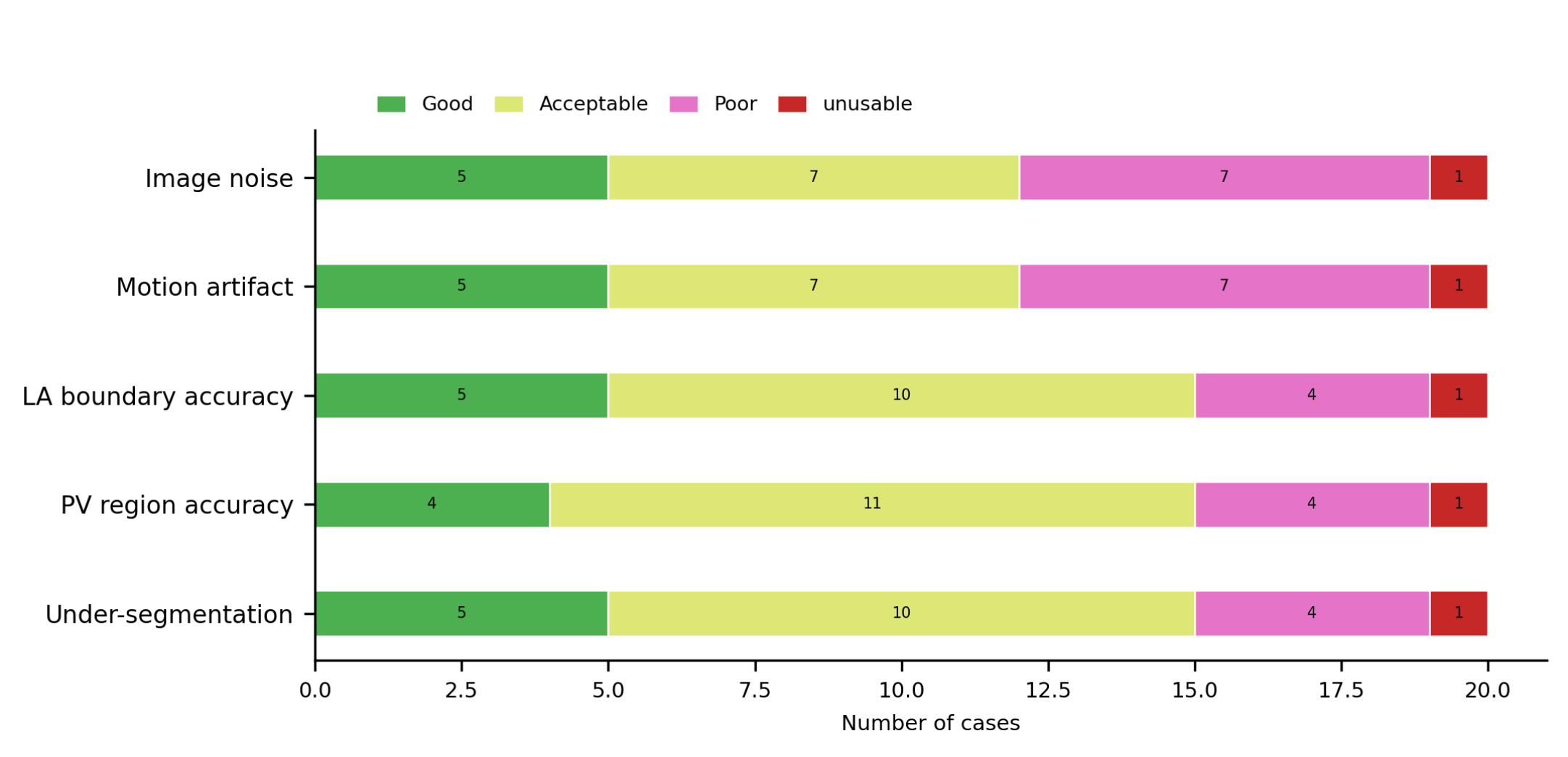}

\caption{Distribution of the radiologist-assigned IQA scores across the five criteria during the construction of the MRI-IQA dataset.} \label{fig1}
\vspace{-1.5em}
\end{figure}

All images were annotated by a single expert radiologists using a pre-defined scoring rubric that reflects the relevant quality attributes. Each case is therefore evaluated across five criteria: Noise, Motion Artifact, LA Boundary Accuracy, Pulmonary Vein (PV) Region Accuracy, and Under-segmentation Severity, along with an overall clinical usability decision for ablation planning. Each criterion is scored on an ordinal scale from 0 (Unusable) to 3 (Good), capturing the graded nature of clinical assessment. Table \ref{tab:criteria_definition} shows the definition of each criteria used for radiologist annotation in this study, while Figure \ref{fig1} shows the distribution of the annotations across the five criteria.

\vspace{-1em}

 \vspace{-1em}
\section{Methodology}

\begin{figure}[t]
\includegraphics[width=\textwidth]{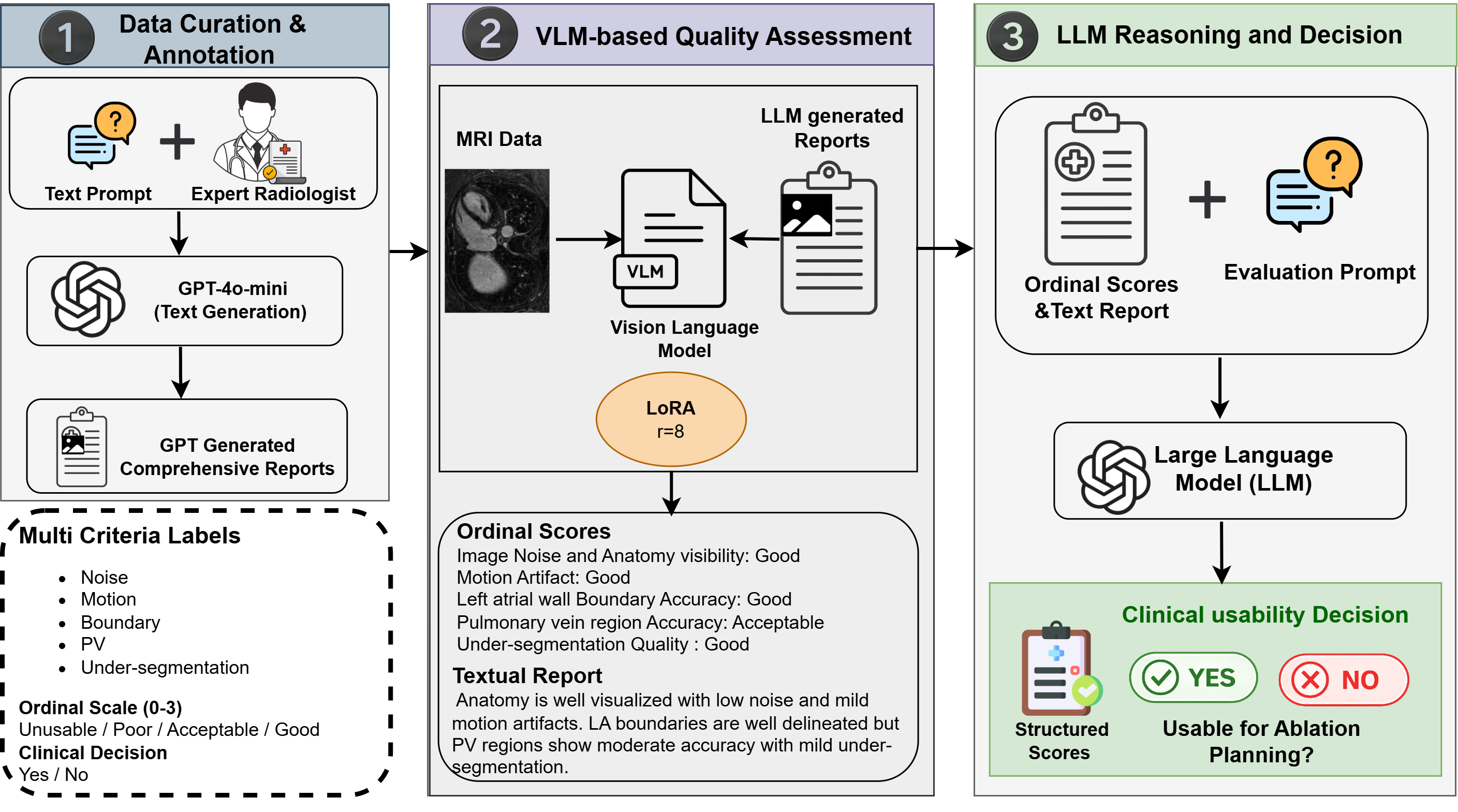}
\caption{Overview of IQA pipeline. Data curation and annotation are followed by LoRA-adapted fine-tuning of the VLM that generates scores and textual reports, which are processed by an LLM to produce a clinical usability decision.} \label{fig2}
\vspace{-1em}
\end{figure}

We formulate our proposed MRI-IQA protocol as a clinically grounded reasoning problem, rather than a direct regression or classification task. In clinical practice, radiologists do not assign scores in isolation. They interpret visual features, form a semantic understanding of image quality, and derive a decision regarding clinical usability. The proposed framework consists of two sequential stages, (1) Image-to-Report Generation, (2) Report-to-Score Mapping, and Clinical Usability Decision. This decomposition enables modular learning of clinically interpretable representations while reducing ambiguity inherent in subjective quality assessment. Figure \ref{fig2} shows the overview of our two-stage framework.

\vspace{-1em}
\subsection{Image-to-Report Generation }
\vspace{-.5em}
Direct score prediction treats IQA as a purely visual regression problem, ignoring the latent clinical reasoning process underlying expert assessment. To address this limitation and standardize the textual supervision, we include an additional quality description along with the image slice-text pair conditioned on expert-provided scores during data curation. We leverage the strength of the LLM to generate the quality description using a controlled prompt template. Specifically, we use the gpt-4o-mini~\cite{achiam2023gpt} API to generate the quality descriptions. This instance is not used at inference and has no role in score extraction. This inclusion acts as a semantic bottleneck, encouraging the model to encode high-level clinical attributes rather than directly optimizing for discrete scores. As a result, using this image slice-text paired dataset, the model learns representations that are more interpretable and better aligned with expert reasoning. After fine-tuning, the VLM generates a structured textual description capturing clinically relevant attributes, including all the criteria mentioned in Table \ref{tab:criteria_definition}.
\vspace{-1.30em}
\subsection{Report-to-Score Mapping and Clinical Usability Decision}
While criterion-level IQA provides detailed characterization of image quality, it does not directly answer the clinically relevant question: Is this image usable for ablation planning. This task aggregates multiple quality attributes into a binary decision reflecting whether an image is suitable for reliable anatomical analysis. The VLM predicted quality and reports are used to generate structured numeric quality scores and derive a binary clinical usability decision based on the given criteria. The reason behind this stage is that it provides robustness to surface-level output variation: a free-text generation that is inherently non-deterministic, plus minor deviations in field naming, punctuation, or phrasing would cause silent failures in a direct parsing approach. On the other hand, the use of a structured JSON schema with strictly constrained output fields eliminates this brittleness. Most importantly, the extraction step performs explicit multi-criteria clinical reasoning to derive the binary usability decision from the quality description of the five predicted attributes. The two-stage design therefore separates perception (VLM) from structured reasoning (LLM), a decomposition that improves both robustness and clinical interpret-ability.

\subsection{Prompt Design}
\subsubsection{Report Generation Prompt:} To generate linguistically varied yet clinically consistent training supervision, we instruct gpt-4o-mini to produce a radiology-style quality assessment comprehensive report conditioned on expert-provided ordinal scores. The prompt supplies five quality scores on a 0–3 scale (0=unusable, 1=poor, 2=acceptable, 3=good) corresponding to Noise, Motion Artifact, LA Boundary Accuracy, PV Region Accuracy, and Under-segmentation severity, and explicitly maps each numeric score to its qualitative descriptor. To prevent the model from producing repetitive, template-like text across training samples that would cause the VLM to memorize surface patterns rather than learn quality-discriminative features, we introduce three per-slice focus variants. Slice 0 is instructed to lead with noise and motion artifact quality, slice 1 with boundary and PV accuracy, and slice 2 with under-segmentation and overall usability. Although all three slices from the same patient share identical ground-truth scores, this variation in linguistic emphasis produces three distinct textual representations per patient, increasing effective training diversity without requiring additional annotations. The model is instructed to use natural radiology-style language and avoid fixed sentence template, as well as not to hallucinate findings that are not supported by the scores and only return the explanation text without labels.
\vspace{-1em}
\subsubsection{Report-to-Score Mapping Prompt} At inference time, predicted quality and free-text reports are converted into structured ordinal scores using a second GPT-based prompt. The system prompt defines gpt-4o-mini as a cardiac MRI QA assistant and provides explicit decision rules for each of the five criteria on a four-point scale (0=unusable, Poor=1, Acceptable=2, Good=3). For the binary clinical usability decision (Yes/ No), the prompt encodes conservative clinical thresholds: an image is classified as usable for ablation planning only if Motion, Boundary Accuracy, and PV Accuracy are all non-Poor. These thresholds were directly derived from the radiologists' annotation patterns associated with this study and reflect anatomically critical criteria for  ablation planning: motion artifacts cause left atrial wall distortion, boundary inaccuracy prevents reliable segmentation, and pulmonary vein region inaccuracy compromises ostia localization. On the other hand, image noise alone does not preclude planning as long as boundaries remain intact, and is therefore excluded as a hard threshold and instead treated as a clinically conservative design that prioritizes patient safety. The model returns a structured JSON object containing five categorical labels, five corresponding numeric scores, and a binary Final Decision. 
\vspace{-1em}
\subsection{Implementation Details}
We experimented with four SOTA VLMs (InternVL2-2B, DeepSeek-v-l-1.3b, Qwen2.5-3B, and LLaVa-1.5-7b) as the base framework for our dataset. We used a 80:20 training : testing data split for all our experiments and trained all of our models using ms-swift framework~\cite{zhao2024swiftascalablelightweightinfrastructure} running on a NVIDIA A100 40 GB GPU. In the training process, we adapted the Rank 8 LORA~\cite{hu2022lora} for fine-tuning the VLM models for 10 epochs where we used the batch size 2,  a gradient accumulation step of 2, and a learning rate of $5 \times 10^{-4}$ and conducted an evaluation on the test set.
\vspace{-1em}
\subsection{Evaluation Metrics}
We evaluate performance using accuracy (ACC) and Pearson Linear Correlation Coefficient (PLCC) for each criterion as the outputs are ordinal scores. For the binary clinical decision task, we report ACC, F1 score, and Cohen's kappa coefficient (k). Given the limited test set size, we additionally report PLCC, as correlation based metrics are more robust to small-sample variance. Cohen's kappa coefficient is included for the natural class imbalance in the binary usability decision in the Yes/No distribution with a limited test set; a model predicting the majority class could achieve high accuracy without meaningful discrimination, and Cohen's Kappa coefficient corrects for this by measuring agreement beyond chance~\cite{landis1977kappa}. Since quality criteria are rated on an ordinal four-point scale, we also report tolerance accuracy (Acc±1) alongside exact-match accuracy motivated by evidence that human inter-observer variability in cardiac MRI quality grading routinely spans one ordinal level~\cite{piccini2020deep}, suggesting that exact-match accuracy alone underestimates clinically acceptable agreement. 
\vspace{-1em}
\section{Result Analysis}
We evaluate the performance of our framework on two tasks: Task 1 evaluates how accurately each VLM predicts the five radiologist-defined ordinal scores; Task 2 evaluates the binary clinical usability decision produced by the GPT-based reasoning module operating on VLM-predicted reports. 

\subsubsection{Task 1. }Table \ref{tab} reports criterion-wise performance for the structured quality report generation across four VLM backbones evaluated on the test set. InternVL2-2B achieves the highest average (Avg) accuracy (ACC=0.65) and PLCC (Avg=0.79), followed by DeepSeek-VL-1.3B (ACC=0.60, PLCC=0.78), demonstrating competitive performance despite its smaller parameter count. Qwen2.5-VL-3B and LLaVA-1.5-7B achieve lower accuracy (0.53 and 0.55, respectively), with correspondingly lower PLCC values, demonstrating a weaker correlation with expert annotations under the fine-tuning configuration used.

Across all models, Noise and Motion criteria consistently performed better than the other criteria. This is consistent with the nature of artifacts because noise and motion  produce intensity disturbances that are detectable from LGE MRI and are therefore easier for VLMs to capture from pixel-level features. Boundary accuracy, PV region accuracy, and under-segmentation show lower agreement across all models  (0.50–0.67 for the stronger models, and 0.42–0.58 for the weaker ones), reflecting the increased anatomical specificity and subjectivity involved in assessing fine structural details of the LA wall and PV ostia.

Notably, tolerance accuracy (Acc ± 1) is substantially higher than exact-match accuracy for all models. InternVL, DeepSeek, and LLaVA achieve perfect tolerance accuracy (1.00) across all criteria, indicating that all predictions fall within one ordinal level of the expert rating. Qwen2.5 achieves an Acc ± 1 of 0.92, implying that the remaining 8\% of the predictions deviate by two levels. Since human inter-observer variability in cardiac MRI quality grading routinely spans one ordinal level, these results suggest that all evaluated models produce clinically acceptable predictions in the large majority of cases, even where exact-match scores are moderate.

\begin{table*}[t]
\centering
\caption{Performance across models for each criterion (Noise, Motion, LA Boundary Accuracy, PV Region Visibility, and Under-segmentation Severity) assessed using Accuracy [ACC~$\uparrow$], Pearson Linear Correlation Coefficient [PLCC~$\uparrow$], and Tolerance Accuracy [Tol\textsubscript{Acc}~$\uparrow$], where $\uparrow$ indicates higher is better.}
\label{tab}
\resizebox{\textwidth}{!}{
\begin{tabular}{l|c|c|c|c|c|c|c}
\hline
Model & \multicolumn{1}{c|}{Metric} & Noise & Motion & Boundary & PV Region & Under-seg & \textbf{Avg}\\
\hline

\multirow{3}{*}{\textbf{InternVL2-2B}} 
& ACC~$\uparrow$  & \textbf{0.75} & \textbf{0.75} & \textbf{0.67} & \textbf{0.50} & \textbf{0.58} & \textbf{0.65} \\
& PLCC~$\uparrow$  & 0.91 & 0.91 & 0.75 & 0.71 & 0.72 & 0.79 \\
& Tol$_{\text{Acc}}$~$\uparrow$ & 1.00 & 1.00 & 1.00 & 1.00 & 1.00 & 1.00 \\
\hline
\multirow{3}{*}{DeepSeek-v-1.3b} 
& ACC~$\uparrow$  & 0.75 & 0.75 & 0.50 & 0.50 & 0.50 & 0.60 \\
& PLCC~$\uparrow$  & 0.91 & 0.91 & 0.71 & 0.71 & 0.71 & 0.78 \\
& Tol$_{\text{Acc}}$~$\uparrow$ & 1.00 & 1.00 & 1.00 & 1.00 & 1.00 & 1.00 \\
\hline
\multirow{3}{*}{LLaVa-1.5-7b} 
& ACC~$\uparrow$ & 0.50 & 0.50 & 0.58 & 0.58 & 0.58 & 0.55 \\
& PLCC~$\uparrow$  & 0.52 & 0.52 & 0.43 & 0.43 & 0.43 & 0.50 \\
& Tol$_{\text{Acc}}$~$\uparrow$ & 1.00 & 1.00 & 1.00 & 1.00 & 1.00 & 1.00 \\
\hline
\multirow{3}{*}{Qwen2.5-3B} 
& ACC~$\uparrow$  & 0.67 & 0.67 & 0.50 & 0.42 & 0.42 & 0.53 \\
& PLCC~$\uparrow$  & 0.64 & 0.64 & 0.50 & 0.48 & 0.50 & 0.55 \\
& Tol$_{\text{Acc}}$~$\uparrow$  & 0.92 & 0.92 & 0.92 & 0.92 & 0.92 & 0.92 \\
\hline

\hline

\end{tabular}
            }
\end{table*}
\subsubsection{Task 2. } Table \ref{tab_2} reports performance on the binary clinical usability decision. Critically, Task 2 operates entirely on the free-text quality scores and reports predicted by the VLM in Task 1, not on ground-truth annotations — meaning any criterion-level prediction error from Stage 1 propagates directly into the usability reasoning. Against this background, DeepSeek's perfect agreement with radiologist decisions  (Acc=1.00, F1=1.00, k=1.00) is particularly noteworthy: it demonstrates that the pipeline can absorb imperfect criterion-level predictions and still arrive at the correct clinical conclusion, which is the property that matters most for real-world deployment. InternVL achieves substantial agreement (Acc=0.92, F1=0.90, k=0.83), followed closely by Qwen2.5 (Acc=0.83, F1=0.80, k=0.67), both indicating a strong agreement with the radiologists' decisions. LLaVa1.5 achieves (Acc=0.75, F1=0.80 and k=0.50) a moderate agreement.

\begin{table}
\centering
\caption{Final clinical usability decision performance across methods.}
\label{tab_2}
\begin{tabular}{lccc}
\toprule
\textbf{Method} & \textbf{Acc~$\uparrow$} & \textbf{F1~$\uparrow$} & \textbf{Cohen's}  \\
                &                         & \textbf{(Y/N)}         & \textbf{Kappa~$\uparrow$} \\
\midrule
\textbf{DeepSeek-v-1.3b} & \textbf{1.00} & \textbf{1.00} & \textbf{1.00}  \\
InternVL2-2B & 0.92 & 0.90 & 0.83  \\

{LLaVa-1.5-7b} & 0.75 & 0.80 & 0.50  \\
Qwen2.5-3B & 0.83 & 0.80 & 0.67  \\
\bottomrule
\end{tabular}
\end{table}

\begin{figure}
\includegraphics[width=\textwidth]{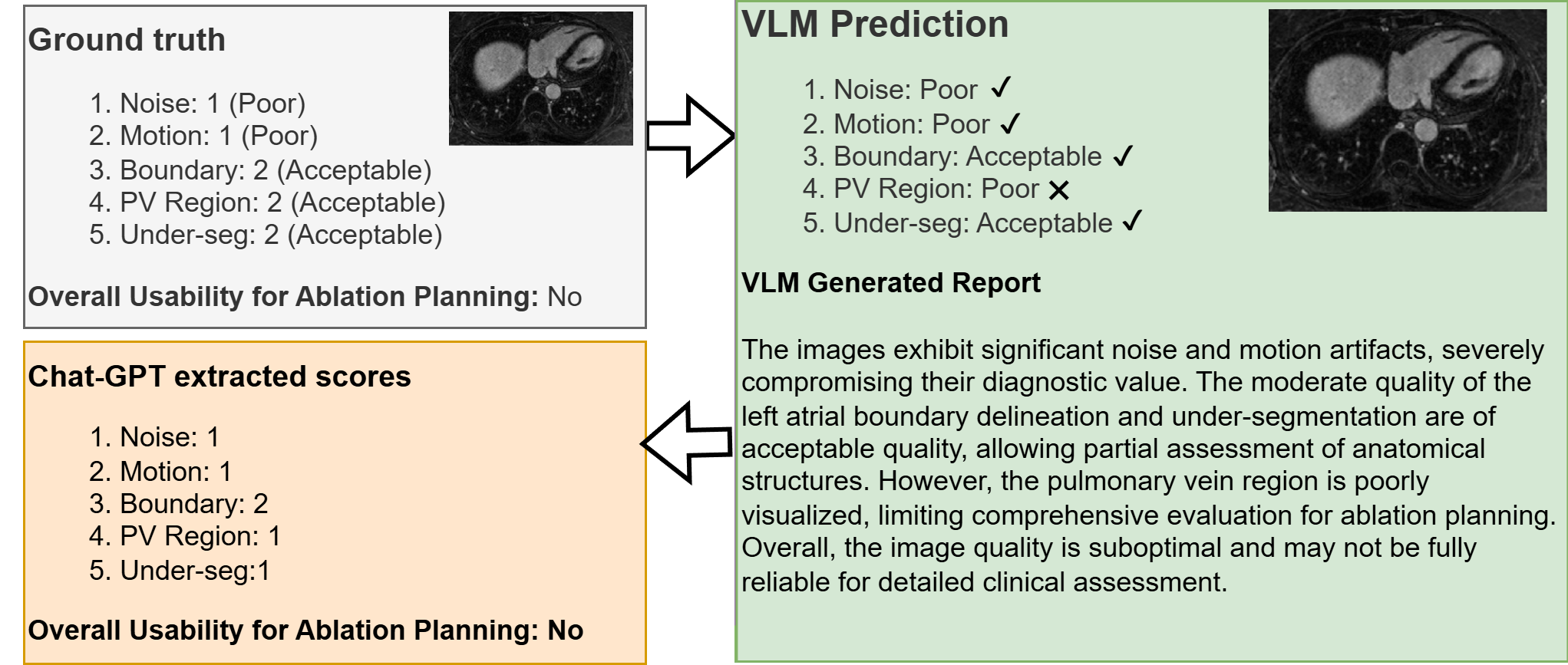}
\caption{Qualitative comparison of VLM Prediction, Chat-GPT extracted scores and clinical decision for usability  with ground truth}
\vspace{-1em}
\label{fig3}
\end{figure}
These results are broadly consistent with theoretical expectations based on architectural design. InternVL's leading performance on ACC and PLCC aligns with its dynamic high-resolution vision encoder, which preserves fine spatial detail critical for assessing thin-walled anatomical structures such as the LA boundary. DeepSeek's strong Task 2 performance, despite its smaller size, is consistent with its hybrid dual-encoder design, which explicitly separates global semantic understanding from local high-resolution feature extraction, a property well-suited to the dual demands of noise assessment and boundary delineation. In contrast, LLaVA's below-expectation performance on boundary and PV criteria reflects the known limitations of CLIP-based vision encoders, which are pretrained on natural image slice-text pairs and lack sensitivity to the fine structural features of cardiac MRI. In Figure \ref{fig3}, we provide a qualitative comparison of predicted output with the ground truth data.

The high performance in predicting overall usability suggests that our proposed framework effectively captures clinically relevant quality patterns and can support decision-making in ablation planning. Notably, even when minor discrepancies occur at the criterion level, the model is often able to produce correct usability decisions, indicating robustness in aggregating quality signals.

\vspace{-01em}

\section{Discussion}
\vspace{-1em}
This study presents a proof-of-concept evaluation of a two-stage VLM-LLM framework for clinically grounded image quality assessment of LGE-MRI. The dataset is limited to 60 image slices from 20 patients which reflects the high cost of expert annotation. Therefore, results should be interpreted as preliminary evidence rather than definitive validation. All annotations were performed by a single expert radiologist using a standardized rubric. While this approach ensures consistency, future work will incorporate multi-reader protocols to quantify inter-rater agreement. 

The dataset is also imbalanced, with very few unusable cases, which indicates the clinical acquisition practice but limits the evaluation of model performance on rare failure cases. Given the limited test set size, PLCC is additionally reported alongside accuracy as a correlation-based metric that is more robust to small-sample variance, and Cohen's kappa is reported for the binary usability decision to account for class imbalance. Moreover, the training descriptions were generated by GPT, conditioned on expert-provided ordinal scores, leveraging the capacity of LLMs to produce linguistically diverse, radiology-style text that mimics clinical reporting patterns~\cite{singhal2023large}. While this introduces a stylistic dependency on the generative model, it provides scalable supervision without requiring additional radiologist annotation effort.

All models except Qwen2.5 achieve perfect tolerance accuracy (Acc±1=1.00), confirming that no model produces a two-level ordinal error that would result in clinically harmful misclassifications. Overall, the results support the feasibility of VLM-LLM based image quality assessment. As part of our future research directions, we aim to include external validation on larger multi-center cohorts before clinical deployment can be considered.

\section{Conclusion}
\vspace{-.7em}
This work presents a two-stage VLM-LLM framework for clinically grounded IQA of left atrial LGE-MR images, where the image quality assessment is formulated as a structured reasoning problem rather than a scalar regression task. A fine-tuned VLM generates radiology-style quality reports capturing five criteria, which are then mapped by a GPT-based reasoning module to a binary clinical usability decision for ablation planning. Evaluated on a dataset of 60 expert-annotated image slices from 20 patients, InternVL achieves the highest criterion-level performance with perfect tolerance accuracy across all criteria, while DeepSeek achieves perfect clinical usability agreement. Based on the data available and the proof-of-concept scale at this stage, the consistency across three of four models suggests that predictions are unlikely to produce harmful two-level misclassifications. The strong Task 2 performance further demonstrates that the report-to-decision pipeline robustly translates structured quality attributes into reliable usability judgments, which remain correct even when individual criterion predictions deviate by one ordinal level. 

These results serve as solid proof-of-concept evidence that speaks for the feasibility of VLM-based clinical image quality assessment under realistic conditions featuring limited annotated data. Therefore, future efforts will focus on the implementation and validation of the proposed model on larger, multi-center cohorts in the effort to confirm generalizability and potential clinical deployability.
\vspace{-1em}

\begin{credits}
\subsubsection{\ackname} We would like to acknowledge the generous support for this work by the National Institutes of Health – National
Institute of General Medical Sciences under Award No.
R35GM128877 and the National Science Foundation - Division of Chemical, Bioengineering and Transport Systems under Award No. 2245152. We also acknowledge the IT help and support with the image data handling provided by Larry Stockmaster, senior analyst in the Information Systems Division at the University of Rochester Medical Center. Lastly, we acknowledge the computational support and access to the computing infrastructure provided by RIT’s Research Computing Division~\cite{https://doi.org/10.34788/0s3g-qd15}.

\subsubsection{\discintname}
The authors have no competing interests to declare that are relevant to the content of this article.
\end{credits}

%
\bibliographystyle{splncs04}
\bibliography{references}

\end{document}